\documentclass[preprint2]{aastex}

\newcommand{\ee}[1]{\mbox{${} \times 10^{#1}$}}

\newcommand{\kms}{\mbox{\,km\,s$^{-1}$}}
\newcommand\cmv{\mbox{\,cm$^{-3}$}}
\newcommand\cmsq{\mbox{\,cm$^{-2}$}}
\newcommand{\um}{\,$\mu$m}
\newcommand{\wno}{\,cm$^{-1}$}

\def\lsim {$\rlap{\raise.4ex\hbox{$<$}}\lower.55ex\hbox{$\sim$}\,$}
\def\gsim {$\rlap{\raise.4ex\hbox{$>$}}\lower.55ex\hbox{$\sim$}\,$}

\def\wig {$\sim$\,}

\newcommand\vibrot{vibration-rotation}
\newcommand{\lsun}{\mbox{\,L$_\odot$}}
\newcommand{\msun}{\mbox{\,M$_\odot$}}

\newcommand{\hh}{H$_2$}

\newcommand{\nhhh}{NH$_3$}
\newcommand{\oco}{CO$_2$}

\newcommand{\hcch}{C$_2$H$_2$}

\newcommand{\chhhh}{CH$_4$}

\newcommand{\nufi}{$\nu_5$}
\newcommand{\nufo}{$\nu_4$}
\newcommand{\nuff}{$\nu_4$+$\nu_5$}
\newcommand{\nuto}{$\nu_2$}

\slugcomment{to be published in ApJ}

\shorttitle{Interpretation of Molecular Spectra}
\shortauthors{Lacy}

\begin{document}


\title{Interpretation of Infrared Vibration-rotation Spectra of
Interstellar and Circumstellar Molecules}


\author{John H. Lacy}
\affil{Department of Astronomy, University of Texas, Austin, TX 78712}
\email{lacy@astro.as.utexas.edu}




\begin{abstract}
Infrared \vibrot\ lines can be valuable probes of interstellar and
circumstellar molecules, especially symmetric molecules, which have
no pure rotational transitions.
But most such observations have been interpreted with an isothermal
absorbing slab model, which leaves out important radiative transfer
and molecular excitation effects.
A more realistic non-LTE and non-isothermal radiative transfer model
has been constructed.
The results of this model are in much better agreement with the
observations, including cases where lines in one branch of a \vibrot\
band are in absorption and another in emission.
In general, conclusions based on the isothermal absorbing slab model can be
very misleading, but the assumption of LTE may not lead to such large errors,
particularly if the radiation field temperature is close to the gas
temperature.
\end{abstract}





\section{The Promise}

Infrared spectroscopy of molecular vibration-rotation lines holds the
promise of providing a valuable probe of chemical and physical conditions
in interstellar and circumstellar gas.
Vibration-rotation spectra complement pure rotational spectra in several
important ways.
First, molecules without permanent dipole moments, such as \chhhh, \hcch,
and \oco, can be observed when vibration breaks the symmetry of the
molecule.
Second, because \vibrot\ lines are typically seen in absorption toward
embedded young stars they selectively probe gas in star-forming regions.
And third, because \vibrot\ lines of different rotational states lie close
together in wavelength they can be observed with one instrument and
telescope, and often simultaneously.

It seems that the interpretation of absorption lines should be
straightforward.
The optical depth of an absorption line depends primarily on the column
density in the lower state, that is ${\rm N_J(v=0)}$, so the calculation of
${\rm N_J}$ from the spectrum is simply a matter dividing the equivalent
width of a line by the line-strength factor obtained from laboratory
spectroscopy.
If many lines are observed, it is not even necessary to assume a
thermal distribution of lower state populations, as the population
of each rotational state can be measured.
Of course, for optically thick lines curve-of-growth or other
corrections for saturation may be necessary.
This introduces uncertainties due to the generally unknown lineshape,
but the problem can be avoided if necessary by observing optically
thin lines of rare isotopes.
This interpretation procedure is referred to as the isothermal
absorbing slab model.
Variations on it have been used by astronomers to interpret almost
all infrared molecular absorption spectra.

\section{The Problems}

There are several problems with the isothermal absorbing slab model:
the dust which emits the `background' continuum radiation is often
mixed with the absorbing gas, the molecular gas emits as well as
absorbs radiation, and there is a temperature gradient in
the gas and dust around embedded sources.

As a first approximation, the effect of mixed gas and dust can be
taken into account by considering the derived molecular column densities
to be the columns to a depth in the source where the dust optical depth,
$\tau$, reaches one.
But since the dust opacity is wavelength dependent, this depth generally
differs for different molecules or different \vibrot\ bands,
and may even differ substantially for lines within a band.
In addition, the emergent continuum radiation at a given wavelength is not
necessarily emitted from near $\tau = 1$.
If there is a temperature gradient in the dust, with the dust in the
source being hotter at greater depth, emission by dust at $\tau > 1$
may dominate the outgoing radiation, especially at wavelengths shortward
of the peak of the Planck function.
Without a radiative transfer calculation that takes these effects
into account, the column of gas through which absorption line observations
are made may be substantially underestimated, and even derived
molecular abundance ratios may be in error.

The effect of emission in the molecular lines is even more difficult
to determine and account for.
If a source is spatially resolved and the molecular gas is in LTE
at a known temperature, the equation of radiative transfer along the
line of sight is easily solved.
The result is that molecular lines saturate at an intensity equal to
the Planck function at the line wavelength.
But often the continuum source is unresolved and the absorbing and
emitting gas may have different angular extents, which may not
fall entirely within the observing beam.
Even more problematical is the fact that collisional deexcitation
cross sections are small \citep{gonzalez02} and radiative rates are
relatively large for \vibrot\ transitions, making the critical densities
for vibrational thermalization large, typically $> 10^{12}$~cm$^{-2}$.
Consequently, vibrational LTE at the gas kinetic temperature could be
a very poor approximation.
It may be a better approximation to assume that all
radiative excitations are followed by radiative decay.
In a spherically symmetric situation in which the molecules lie in
a shell separated from the continuum source, but within the observing
beam, there may be no net absorption even if the lines are optically
thick, since an equal number of photons are emitted into the beam as
are absorbed out of the line of sight.
The actual net absorption or emission depends on whether more or less
molecular gas lies along the observing line of sight than along a
typical direction and whether reemitted photons are absorbed by
dust before escaping from the source region.
In the case of sources with systematic motions, notably expanding
shells around evolved stars, the emission and absorption may be
separated spectrally by the Doppler effect.
In those cases high spectral resolution observations would avoid
this problem, but in many other cases the systematic motions are smaller
than turbulent linewidths or are unresolved, making it impossible to
know how much the emission and absorption cancel.

In many cases the radiative transfer effects we have been discussing
change the strengths of lines, so lead to errors in derived abundances,
but are not obvious from inspection of the spectra.
But in a few cases there are recognizable symptoms of these effects.
The importance of emission in \vibrot\ lines is clearest in the spectra
of expanding circumstellar shells, with IRC+10216 being a particularly
good example.
The bending-mode Q branches of HCN and \hcch\ near 14\um\ are prominent
in the R=2000 ISO SWS spectrum of IRC+10216 \citep{cernicharo99}, with
depths of \wig20\% of the continuum, but the P and R branch regions
of the spectrum show little evidence of line absorption or depression
of the continuum due to lines.
In contrast, the R=80,000 TEXES observations of \citet{fonfria08} 
show the R branch lines of HCN and \hcch\ to have typical depths of
\wig50\%.
The reason the lines are so much weaker in the ISO SWS spectrum is that
the lines have P-Cygni profiles, with nearly equal emission and 
absorption components that are blended together in the R=2000 spectrum.
Apparently most vibrational excitations caused by absorption of
photons in these bands are followed by emission, and most emitted
photons escape from the shell, perhaps after multiple absorption
and emission events.

\begin{figure}
\includegraphics[angle=0,scale=0.5,clip=true]{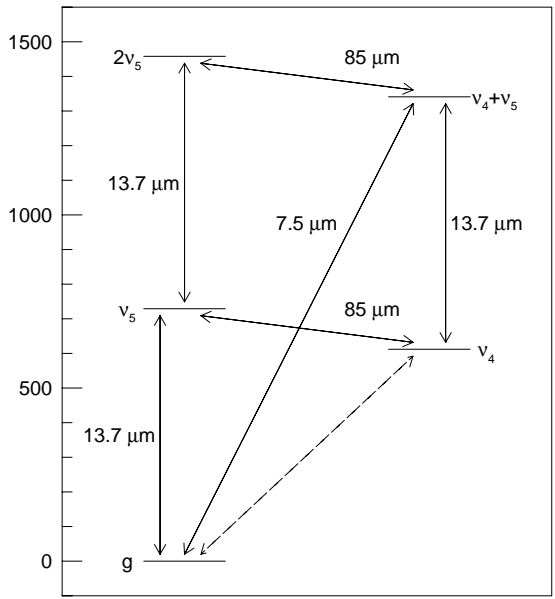}
\caption{\hcch\ vibrational energy levels and transitions of interest.
The \nufo\ and \nufi\ modes are the anti-symmetric and symmetric bending
modes, respectively.
Allowed radiative transitions are indicated with solid arrows;
the \nufo\ transition that only occurs collisionally is indicated
with a dashed arrow.
Allowed radiative transitions between these levels must change the
quantum number v$_5$; v$_4$ can change only if v$_5$ also changes.
The 85\um\ $\nu_5 - \nu_4$ transitions have not been observed.
Rotational levels and level splitting are not shown.
The vertical scale is in \wno.}
\end{figure}

The 13.7\um\ \nufi\ band of \hcch\ has also been observed
toward a number of embedded high-mass young stars
\citep{lacy89, evans91, lahuis00, knez09, barentine12}.
\citet{evans91} and \citet{barentine12} also observed absorption
in the 7.5\um\ \nuff\ band toward OMC1 IRc2 and NGC\,7538 IRS\,9.
An energy level diagram of \hcch\ with the relevant vibrational
levels is shown in Figure 1.
Both bands involve absorption from the ground vibrational level,
but the derived \hcch\ column density from the \nuff\ lines is
greater than that derived from the \nufi\ lines by a factor of
3-10 toward IRc2 and 20 toward IRS\,9.
Both of the effects described above may contribute to this discrepancy.
The dust opacity at 13.7\um\ is about twice as large as that
at 7.5\um, because of the wings of the 9.7 and 18\um\ silicate
features.
In addition, 7.5\um\ is farther off of the peak of the dust thermal
emission, also causing the continuum to be formed farther into the
sources (where the temperature is higher) at 7.5\um.
The amount of reemission in spectral lines may also differ between
the \nufi\ and the \nuff\ bands, since the \nufi\ state can only
decay to the ground via the \nufi\ band, whereas the \nuff\ state
can decay either to the ground through emission of a \nuff\ photon or 
to the \nufo\ level by emission of a $\nu_4$+$\nu_5$-$\nu_4$ photon.

Ammonia (\nhhh) also shows evidence of radiative transfer effects.
\citet{barentine12} observed lines in two branches of the \nhhh\
\nuto\ band toward NGC\,7538 IRS\,9.
Although other molecules seen toward IRS\,9, including \hcch\ and HCN,
are seen in absorption and the aQ branch of \nhhh\ shows P-Cygni lines
or weak absorption, aP and sP-branch \nhhh\ lines are seen in emission.
\citet{barentine12} briefly discuss a radiative transfer
model to explain these observations.
It assumes that the \nhhh\ molecules are exposed to optically thin
silicate emission, which is stronger on the \nhhh\ R and Q branches,
so causes more upward transitions in those branches than in the P branch.
Since downward transitions are about equally likely in the three branches,
emission dominates over absorption in the P branch, and
absorption dominates over emission in the Q and R branches.

\section{A Radiative Transfer Model}

To reproduce and understand the radiative transfer and radiative
excitation effects in molecular \vibrot\ spectra, I constructed a
computer model with includes collisional and radiative excitation and
deexcitation of molecules and calculates the outgoing spectrum.

The model assumes a parameterized distribution and composition of gas
and dust in a shell around a blackbody luminosity source.
The input parameters can be adjusted through a grid search procedure
to fit the continuum and line spectra of individual sources.
For the calculations in this paper we assume a spherically symmetric
expanding shell.
The model first calculates the dust temperature and continuum radiation
field by alternately solving the equation of radiative transfer and
the equation of thermal equilibrium for the dust.
The radiative transfer calculation involves following a grid of rays
through the shell.
Typically, a 64$\times$64 ray grid is used at the outer edge of the model.
After following the rays half way in to the star, the inner 32$\times$32
rays are split into 64$\times$64 rays to increase the resolution.
This is done up to 11 times, and then the rays are rejoined going out
from the star.
Rays that overlap the star are replaced by an appropriate fraction of
the stellar spectrum before continuing out through the shell.
As the rays are followed through the shell, the ray spectrum is added
to a mean intensity spectrum stored for each grid cell, which is used
to calculate the dust heating.
The dust temperatures are calculated for cells on a spherical coordinate grid.
For each grain type and size the equilibrium temperature is calculated
that balances radiative heating and cooling.
The dust source function is then recalculated and the radiative transfer
calculation repeated.
Typically 16 iterations of the radiative transfer and thermal equilibrium
calculations are sufficient for even very optically thick dust shell models
to converge.
The gas temperature is not calculated separately, but is assumed to
equal the dust temperature.
For the models described in this paper, the dust shell and stellar
parameters were chosen to roughly reproduce the observed SED of
NGC\,7538 IRS\,9.

After solving for the shell thermal structure, the program solves
for the molecular level populations and the outgoing line spectrum.
This is done by alternately solving the equation of radiative transfer
in molecular lines and the equation of statistical equilibrium involving
collisional and radiative rates in and out of vibrationally excited
energy levels.
Line strengths are calculated from band strengths and H\"onl-London
factors, as in \citet{evans91}, and are consistent with those in the
HITRAN database.
The calculation is simplified for linear and symmetric top molecules
by the selection rules that cause each vibrationally excited rotational
level to be coupled radiatively to only two or three rotational levels
in the ground vibrational state.
It is further simplified by assuming that the rotational levels in the
ground vibrational state are in LTE at the gas temperature.
This and other assumptions are discussed in \S 5.
We also assume that collisional transitions between vibrational states
do not change the rotational state.
This is not actually valid, but since collisions must drive populations
toward LTE, and the ground vibrational state is assumed to be in LTE,
on average collisions should not change the rotational excitation
of molecules.
In addition, vibration-changing collisional transitions are relatively
unimportant compared to radiative transitions at the densities found in
the model.
With these simplifications, the excited \vibrot\ level populations
and the \vibrot\ lines between them and the ground vibrational state
can be calculated for one excited \vibrot\ level at a time, neglecting
any coupling between rotational levels of the excited vibrational state.

As with the dust calculation, the code follows a grid of rays through the shell,
in this case calculating the spectrum for wavelengths close to the relevant
\vibrot\ lines.
At each step through the shell the molecular line profile is
Doppler shifted by the component of any assumed systematic motion
along the ray direction.
As the rays propagate through the shell the Doppler-shifted spectrum
is added to a sum for the rays passing through each spherical
coordinate cell, so that the mean intensity, averaged over each
spectral line, can be calculated for each cell.
With the mean intensities, the radiative excitation and deexcitation rates
are calculated, and with those and the collisional rates,
the \vibrot\ level populations are calculated.
The radiative transfer and level population calculations are then
repeated until they converge.
However, the fact that the molecular lines are typically more optically
thick than the dust continuum results in slower convergence than
for the calculation of the thermal structure.
This problem was overcome by comparing the change in level populations
in each iteration with that in the previous iteration, and amplifying
the correction if iterations are well correlated.
With the accelerated convergence scheme, the level populations almost
always converge within 16 iterations.

\section{Results of Modeling}

Models were run with parameters meant to roughly reproduce the IRS 9
observations.
The central source was taken to be a 12,000 K, $10^4$\lsun\ blackbody.  
The gas and dust shell had a constant \hh +He density of 2\ee{7}\cmv\ from
4 to 400 AU,
then fell with an $r^{-2}$ density profile out to 1.6\ee{4} AU.
Its mass was 4.2\msun, and its radial column density was 1.8\ee{23}\cmsq.
The calculated dust temperature was 1800\,K at the inner edge of the shell,
falling to 160\,K at 400\,AU and 40\,K at the outer edge of the shell.
A constant expansion speed of 2\kms\ and a turbulent velocity dispersion
of 3\kms\ were assumed.
Molecular line profiles were calculated for the \hcch\ \nufi\ and \nuff\
bands, with \hcch /\hh\ = 1\ee{-7}, the \nhhh\ \nuto\ band, with
\nhhh /\hh\ = 4\ee{-6}, and the CO v=1-0 band, with $^{13}$CO/\hh\ = 1\ee{-6}
and C$^{18}$O/\hh\ = 2\ee{-7}.
These parameters are not meant to be fitted values, but they
produced line depths similar to those observed.

\begin{figure}
\includegraphics[angle=0,scale=0.45,clip=true]{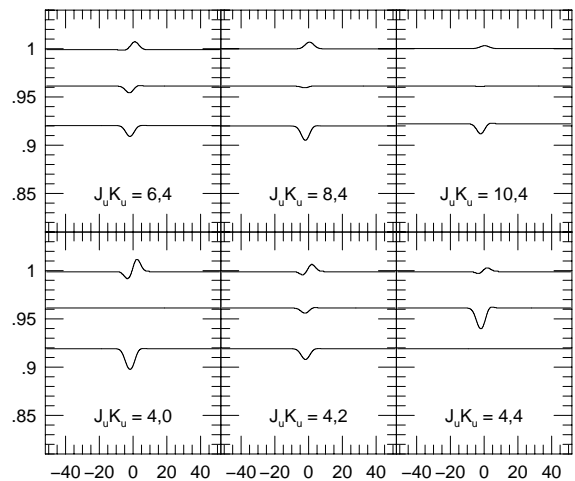}
\caption{Model \nhhh\ \nuto\ umbrella-mode line profiles.
In each panel, the P, Q, and R-branch lines between a given J,K state
in the excited vibrational level and the ground are shown (top to bottom).
Spectra are normalized, with the Q and R-branch lines offset downward.
Q-branch lines are strongest for K$_u$ = J$_u$ and are absent for K$_u$ = 0.
R-branch lines are absent for K$_u$ = J$_u$.}
\end{figure}

Several of the calculated \nhhh\ line profiles are shown in Figure 2.
In each panel, the P, Q, and R-branch lines between a given J,K state
in the excited vibrational level and the ground vibrational level are shown.
Note that these are normally referred to as the P(J$_u$+1,K), Q(J$_u$,K),
and R(J$_u$-1,K) lines, being labeled by their rotational levels
in the lower vibrational state.
For this `umbrella' vibrational mode, Q-branch lines have the largest
optical depths for K=J, and P and R-branch lines are strongest for K=0.
The nuclear statistical weights favor lines with K=3n by a factor of 2.
In all cases, the model P-branch lines are predominantly in emission,
the Q-branch lines show weak absorption, and the R-branch lines are 
more strongly in absorption.
This is in agreement with the observed P and Q-branch lines toward IRS\,9.
The R-branch lines have not been observed, but are predicted to be seen
in absorption, with depths of 1-2\%.

Several calculated \hcch\ \nufi\ lines are shown in Figure 3,
and \nuff\ lines are shown in Figure 4.
For \hcch, rotational states in the ground vibrational level with
J$_l$ odd (ortho states) have nuclear statistical weights of 3;
those with J$_l$ even (para states)
have nuclear statistical weights of 1.
Line-strength factors for P, Q, and R-branch lines are proportional to
J$_u$, 2J$_u$+1, and J$_u$+1, respectively.
Combining these factors for the lines shown, all of which have J$_u$ even,
so J$_l$ odd for the P and R-branch lines, the P and R-branch lines
have similar line-strength factors, whereas Q-branch lines are weaker
on average by a factor of 2/3.
In addition, for absorption lines the line-strength factors must be
multiplied by the Boltzmann population factors of the lower rotational
states.

\begin{figure}
\includegraphics[angle=0,scale=0.55,clip=true]{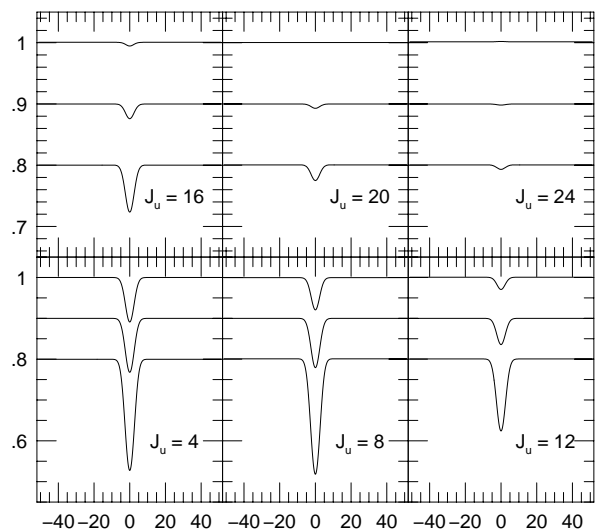}
\caption{Model \hcch\ \nufi\ symmetric bending mode line profiles.
In each panel, the P, Q, and R-branch lines between a given J$_u$ level
in the excited vibrational level and the ground are shown (top to bottom).
Q and R-branch lines are offset downward.
Rotational levels of the excited vibrational level are 
split into ortho and para states, with the P and R-branch lines going
to the ortho (g$_n$ = 3) state for J$_u$ even, and the Q-branch line
going to the para (g$_n$ = 1) state.}
\end{figure}

\begin{figure}
\includegraphics[angle=0,scale=0.45,clip=true]{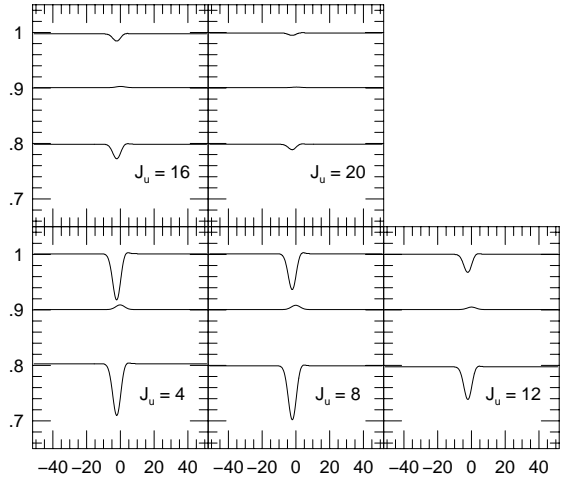}
\caption{Model \hcch\ \nuff\ band line profiles.
Only P-branch (top) and R-branch (bottom) lines occur for this band;
the middle spectrum is for the $\nu_4+\nu_5-\nu_4$ Q-branch line,
which accounts for much of the radiative deexcitation of the \nuff\ state.
Since it is scaled by the continuum flux, the $\nu_4+\nu_5-\nu_4$ line
appears weaker than the \nuff\ lines.}
\end{figure}

In agreement with the IRS\,9 observations, the \nufi\ R-branch lines have
depths only about twice those of the \nuff\ lines, even though the
\nufi\ band strength is nearly 10 times greater.
The P-branch line depths are similar in the two bands.
A new prediction is that the \nufi\ P-branch lines should be
substantially weaker than the R-branch lines.
At higher abundances the P-branch lines go into emission and the
Q-branch lines show P-Cygni profiles.
The P-branch lines are unobservable from the ground, due to telluric
\oco\ absorption.
The Q-branch lines are difficult to observe, due to telluric absorption
and blending of the closely spaced lines, but do show some evidence
of P-Cygni emission.
The explanation of the strengths and profiles of these various lines
is discussed below.

\begin{figure}
\includegraphics[angle=0,scale=0.45,clip=true]{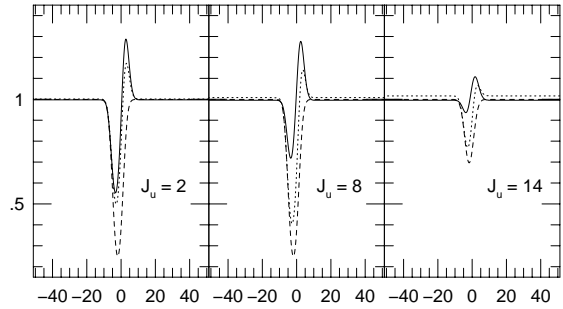}
\caption{Model $^{13}$CO v=1-0 line profiles.
Only P (solid) and R-branch (dotted) lines occur for $^1\Sigma$ diatomic
molecules like CO.
The dashed line in each panel is for an average of the P and R-branch
line strengths, but leaving out emission in the line.}
\end{figure}

Calculated CO lines are shown in Figure 5.
CO has no Q branch; the third line shown in each panel is for an
average of the P and R-branch line strengths, but with emission by
molecules in the v=1 state turned off in the program, to show what
would be observed if the pure-absorption model used in the past
were valid.
Both P and R-branch lines of CO have P-Cygni profiles, with the
relative strength of the emission component being greater in the
P-branch lines.
The observed lines of CO from IRS\,9 also have P-Cygni profiles,
but are considerably broader than the lines of other molecules.
This is probably a result of a high velocity outflow with a different
chemistry from the gas that dominates the spectra of other molecules.
The high optical depths of the $^{12}$CO and $^{13}$CO lines may
also contribute to the prominence of line wings that are not apparent
in spectra of other molecules.
C$^{17}$O and C$^{18}$O lines are blended with $^{12}$CO and $^{13}$CO lines
in the existing observations; further observations would be desirable.

\section{Explanation and Discussion}

The model spectra are generally in at least qualitative agreement
with the observations of IRS\,9.
In particular, they reproduce the emission seen in \nhhh\ P-branch lines
and the similar depths of \hcch\ \nufi\ and \nuff\ lines.
However, they make additional predictions not anticipated by the
simple radiative transfer models of \citet{barentine12}.
Notably, they predict that \hcch\ should show differences between its
P, Q, and R-branch lines like those seen in \nhhh.
Somewhat less prominent effects are predicted for CO.
The explanation given by \citet{barentine12} for \nhhh\ was based
on the brighter continuum radiation at the Q and R-branch wavelengths.
But this should not be the case for \hcch, for which the dust opacity
is similar at the wavelengths of its different branches.

To help understand the radiative transfer effects responsible for the
line profiles of different lines, several physical effects were changed
in the model.
First, a model was run assuming LTE populations of the excited \vibrot\
levels.
Perhaps surprisingly, the model results did not change greatly;
P-branch lines still favored emission relative to R-branch lines.
Apparently the non-LTE populations were not very different from LTE
populations.
The explanation for this may be that the continuum radiation field
seen by the molecules is not very different from a blackbody field
at the dust temperature, and the gas temperature was assumed to be
equal to the dust temperature.
As a result, populations controlled by radiation were not very different
from populations controlled by collisions.
This would not be the case if the molecular gas were more distant
from the warm dust responsible for the continuum, so the conclusion
that non-LTE effects were relatively small for the IRS\,9 model
may not be valid in general.
A second modification to the physics of the model was to set the
rotational constant (B$_e$) of the molecules to zero, so that the
P, Q, and R-branch lines all fell at the same wavelength, and their
lower state energies were all the same.
The differences between P, Q, and R-branch lines then disappeared.
Finally, a model was run with the gas and dust temperature held constant
throughout the shell.
Again, the differences between P, Q, and R-branch lines largely disappeared.
The conclusion from these models with modified physics was
that the difference between the lower state energies,
and the associated Boltzmann factors, was the primary cause of the
different line profiles.

With the results of the modified models, the explanation for the
differences between the P, Q, and R-branch lines can be understood.
We will consider P and R-branch lines, which have similar behaviors
for all molecules.  (The strengths of Q-branch lines depend on the
molecular symmetry and vibrational mode.)
The P and R-branch lines from a given rotational state of the upper
vibrational level have nearly equal Einstein A coefficients
(depending weakly on J$_u$), so nearly equal branching ratios on
emission.
However, the absorption in these lines depends on the radiation field
and the populations of the rotational states
of the lower vibrational level.
Especially for high values of J, these populations can differ
substantially, because of the different energy levels of $J_l = J_u+1$
and $J_l = J_u-1$ for the P and R-branch lines, respectively.
As a result, the two lines have similar emission strengths, but the
R-branch line has greater absorption.
If most upward transitions are followed by a downward radiative transition,
emission and absorption nearly cancel, but with emission dominating
in the P-branch line and absorption dominating in the R-branch line.
In the case of \nhhh\ the greater radiation field at the wavelengths of
the R-branch lines enhances the effect, but the effect is present even for
\hcch\ and CO in models with a temperature gradient through the shell.

The disappearance of the difference between P and R-branch lines
in the isothermal model can be 
explained by the fact that the continuum radiation field at the
R-branch wavelengths is weaker than that at the P-branch wavelengths
for a given J$_u$ by the same Boltzmann factor that the lower level
population is greater, resulting in approximately the same number of
upward transitions in the two lines.
The temperature gradient in the dust shell model caused the color
temperature of the continuum radiation field to be higher than the
gas temperature where the lines are formed.
As a result, heat must flow from the radiation field into the gas.
This happens by net absorption of R-branch photons and net emission
of P-branch photons, pumping the rotational populations.

It would be very desirable to estimate the magnitude of the error that
is made using the isothermal absorbing slab model.
For the models shown here, R-branch lines are typically a factor \wig 2
weaker when reemission is included, but P-branch lines may be 
dominated by emission.
Unfortunately, the extent to which emission cancels absorption depends
on the probability that an emitted photon escapes, rather than being
absorbed by dust, and the asymmetry of the source, which determines
how the absorption along our line of sight compares to that along
other lines of sight.
An understanding of the geometry of the source being observed would be
needed to make a realistic model including reemission.

Several simplifications and approximations made in the model should be noted.
First, the gas temperature was assumed to be equal to the dust temperature.
This assumption may be valid through much of the modeled shell,
even if the density is too low for collisions with dust grains to dominate
the heating and cooling of the gas, since the radiation field temperature
is close to the dust temperature through most of the shell.
However, close to the exciting star ultraviolet radiation may heat the
gas to a higher temperature than the dust, via the photoelectric effect.
The optical depth of the dust in the model is high enough to hide this
region, but in the case of less deeply embedded objects, where the
gas near the exciting source can be observed, departure of the gas
temperature from the dust temperature could be important.
To test this effect, a less optically thick model was run that
included photoelectric heating
of the gas and allowed the gas and dust temperature to differ.
In this model, the gas temperature was greater than the infrared radiation
temperature, so energy should flow from the gas into the radiation.
In fact, P-branch lines then showed greater absorption than R-branch lines,
as is expected by this thermodynamic argument.

Another assumption was that the rotational temperature of the molecules was
equal to the gas temperature.
This might be justified by the fact that the density in most of the
shell is greater than the critical density for rotational thermalization,
especially for symmetric molecules, like \hcch\ and \chhhh ,
which have no allowed rotational transitions.
However, the infrared pumping that led to differing P and R-branch line
profiles may modify the populations.
The importance of infrared pumping can be estimated by comparing the
rate of collisional transitions between rotational levels to the rate
of infrared transitions between the ground and excited vibrational levels.
The collisional transition rate is given by the product of the collisional
rate coefficient, which is typically $\sim 10^{-10}$, and the gas density,
giving a rate $\sim 2 \times 10^{-3}$s$^{-1}$.
The radiative rate is given by the product of the vibrational A coefficient
and the fractional population of the excited vibrational state.
For \hcch , with an A coefficient of 6 s$^{-1}$ for the sum of the P and
R-branch lines to a given rotational state, infrared pumping
should dominate for vibrational temperatures $> 130$ K, or throughout
much of the model shell.
This calculation may overestimate the importance of infrared pumping
since absorption followed by emission of a \vibrot\ photon can at most
change the rotational quantum number by 2, whereas large $\Delta$J can
occur in collisional transitions.
But if the rotational populations are controlled by infrared pumping,
the rotational temperature should equal the infrared radiation color
temperature, and
the difference between P and R-branch lines should disappear.
The resolution of this problem may be that the density of the gas
around IRS\,9 is enough greater than that assumed in the model so
that collisions, rather than infrared pumping control the rotational
level populations.
A similar effect is the pumping of vibrational levels by radiative and
collisional transitions to other vibrational states.
For example, the \nufi\ state of \hcch\ could be populated by collisional
or radiative transitions from the \nufo\ state, which itself has several
possible methods of excitation.
To include the various neglected ways in which the excited \vibrot\
states can be populated would require the simultaneous calculation
of many \vibrot\ states.
Unfortunately, that is beyond the capability of the model used here.

\section{Conclusions}

The first conclusion of this work is that interpretation of \vibrot\
spectra based on the isothermal absorbing slab model can be very
misleading.
In general, column densities derived with this model can be expected
to be underestimates, due to the neglect of reemission by molecules
following absorption, but the magnitude of the error depends on the
source geometry and the lines observed.

The second conclusion is more positive.
Although collisions are not normally sufficiently frequent to maintain
LTE at the gas kinetic temperature, the radiation field temperature
may be similar to the gas temperature, and so molecular level
populations may nevertheless be close to LTE populations.
This was the case for the model considered, and should generally be
the case if the dust optical depth is large.
As a result, non-LTE effects may not be large.
Effects of radiative transfer through gas and dust with a temperature
gradient are likely to be more important.
Fortunately, these effects are easier to include in models used to
interpret observations, and they should be included.

\acknowledgments
I thank the anonymous referee for several interesting comments,
in particular pointing out the possibility of infrared pumping of
the rotational populations.

\end{document}